\documentclass[twocolumn]{emulateapj-rtx4}

\usepackage[]{hyperref}
\hypersetup
{
  bookmarks=true,%
  unicode=false,%
  pdftoolbar=true,
  pdfmenubar=true,
  pdffitwindow=false,
  pdfstartview=FitH,
  pdftitle={Astrophys. J.},
  pdfauthor={LAM Yi Hua},
  pdfsubject={Nuclear Astrophysics},
  pdfcreator={LAM Yi Hua},
  pdfproducer={LAM Yi Hua},
  pdfkeywords={Shell Model},		
  bookmarksnumbered,%
  colorlinks=true,%
  linkcolor=blue,%
  citecolor=blue,%
  filecolor=green,%
  urlcolor=blue,
  breaklinks, 
  plainpages=false%
}

\shorttitle{Reaction rates of $^{64}$Ge($p$,$\gamma$)$^{65}$As($p$,$\gamma$)$^{66}$Se ...}
\shortauthors{Y.H. Lam et al.}

\usepackage{xcolor}
\begin{document}


\title{Reaction rates of $^{64}$Ge($p$,$\gamma$)$^{65}$As and $^{65}$As($p$,$\gamma$)$^{66}$Se and the extent of nucleosynthesis in type I X-ray bursts}


\author{Y.H. Lam\altaffilmark{1}, J.J. He\altaffilmark{1}, A. Parikh\altaffilmark{2,3}, H. Schatz\altaffilmark{4}, B.A. Brown\altaffilmark{4} \\
M. Wang\altaffilmark{1}, B. Guo\altaffilmark{5}, Y.H. Zhang\altaffilmark{1}, X.H. Zhou\altaffilmark{1}, H.S. Xu\altaffilmark{1}}
\affil{\altaffilmark{1}Key Laboratory of High Precision Nuclear Spectroscopy, Institute of Modern Physics, Chinese Academy of Sciences, Lanzhou 730000, China}
\affil{\altaffilmark{2}Departament de F\'{\i}sica i Enginyeria Nuclear, EUETIB, Universitat Polit\`{e}cnica de Catalunya, Barcelona E-08036, Spain}
\affil{\altaffilmark{3}Institut d'Estudis Espacials de Catalunya, Barcelona E-08034, Spain}
\affil{\altaffilmark{4}Department of Physics and Astronomy and National Superconducting Cyclotron Laboratory, Michigan State University, East Lansing, Michigan 48824-1321, USA}
\affil{\altaffilmark{5}China Institute of Atomic Energy, P. O. Box 275(10), Beijing 102413, China}

\email{jianjunhe@impcas.ac.cn; anuj.r.parikh@upc.edu; schatz@nscl.msu.edu}

\begin{abstract}
The extent of nucleosynthesis in models of type I X-ray bursts and the associated impact on the energy released in these explosive events are sensitive to
nuclear masses and reaction rates around the $^{64}$Ge waiting point. Using the well known mass of $^{64}$Ge, the recently measured $^{65}$As mass, and
large-scale shell model calculations, we have determined new thermonuclear rates of the $^{64}$Ge($p$,$\gamma$)$^{65}$As and $^{65}$As($p$,$\gamma$)$^{66}$Se
reactions with reliable uncertainties. The new reaction rates differ significantly from previously published rates. Using the new data we analyze
the impact of the new rates and the remaining nuclear physics uncertainties on the $^{64}$Ge waiting point in a number of representative one-zone X-ray
burst models. We find that in contrast to previous work, when all relevant uncertainties are considered, a strong $^{64}$Ge $rp$-process waiting point cannot
be ruled out. The nuclear physics uncertainties strongly affect X-ray burst model predictions of the synthesis of $^{64}$Zn, the synthesis of nuclei beyond
$A=64$, energy generation, and burst light curve. We also identify key nuclear uncertainties that need to be addressed to determine the role of the $^{64}$Ge
waiting point in X-ray bursts. These include the remaining uncertainty in the $^{65}$As mass, the uncertainty of the $^{66}$Se mass, and the remaining
uncertainty in the $^{65}$As($p$,$\gamma$)$^{66}$Se reaction rate, which mainly originates from uncertain resonance energies.
\end{abstract}


\section{Introduction}
A type I X-ray burst (XRB) arises from a thermonuclear runaway in the accreted envelope of a neutron star in a close binary star system (for reviews, see,
e.g.,~\citet{lew93,sch98,str06,par13}). Roughly 100 bursting systems have been discovered to date, with light curves exhibiting peak luminosities of
$L_\mathrm{peak}\approx$~10$^{4}$--10$^{5}$ $L_\mathrm{\odot}$ and timescales of 10--100 s. During an XRB, models predict that a H/He-rich accreted envelope
may become strongly enriched in heavier nuclei though the $\alpha$$p$-process and the $rp$-process~\citep{wal81,sch98}. These two processes involve
$\alpha$-particle-induced or proton-capture reactions on stable and radioactive nuclei, interrupted by occasional $\beta$-decays. When the $rp$-process
approaches the proton dripline, successive capture of protons by nuclei is inhibited by a strong reverse photodisintegration reaction rate. The competition
between the rate of proton capture and the rate of $\beta$-decay at these ``waiting points" (e.g., $^{60}$Zn, $^{64}$Ge, and $^{68}$Se) determines the extent
of the synthesis of heavier mass nuclei during the burst~\citep{sch98}. Peak temperatures during the thermonuclear runaway may approach or
exceed 1~GK, resulting in the synthesis of nuclei up to mass $A\approx100$~\citep{sch01,elo09}. Model predictions depend, however, on
astrophysical parameters such as accretion rate, the composition of the accreted material, and the neutron star surface gravity, as well as on nuclear physics
quantities such as nuclear masses and reaction rates.

The $^{64}$Ge($p$,$\gamma$)$^{65}$As and $^{65}$As($p$,$\gamma$)$^{66}$Se reactions have been demonstrated to have a significant impact on
nucleosynthesis during XRBs. (See~\citet{par14} for a recent review of the impact of nuclear physics uncertainties on predicted yields and light curves from
XRB models.) Direct measurements of these reactions at the relevant energies in XRBs are not yet possible due to the lack of sufficiently intense radioactive
$^{64}$Ge and $^{65}$As beams. Moreover, due to the unknown mass of $^{66}$Se and the lack of nuclear structure information for states within $\approx$~1--2~MeV
of the $^{64}$Ge+$p$ and the (theoretical) $^{65}$As+$p$ energy thresholds in $^{65}$As and $^{66}$Se, respectively, it is not possible to estimate rates for
these reactions based on experimental nuclear structure data. As a result, XRB models use $^{64}$Ge($p$,$\gamma$)$^{65}$As and
$^{65}$As($p$,$\gamma$)$^{66}$Se thermonuclear rates derived from theoretical calculations. Using such models it has been demonstrated that varying
the $^{65}$As($p$,$\gamma$)$^{66}$Se rate by a factor of ten at the relevant temperatures affects the calculated abundances of nuclei between $A\approx$~65--100
by factors as large as about 5~\citep{par08}. For $^{64}$Ge($p$,$\gamma$)$^{65}$As, models have illustrated the importance of the $Q$-value (or proton separation
energy $S_p$) adopted for this reaction, with variations by $\pm$300~keV affecting final calculated abundances between $A\approx$~65--100 by factors as large
as about 5~\citep{par08,par09}. In addition, the effective $rp$-process lifetime of the waiting-point nucleus $^{64}$Ge was investigated by~\citet{sch066}
based on the estimated proton separation energies of $S_p$($^{65}$As)=-0.36$\pm$0.15 MeV and $S_p$($^{66}$Se)=2.43$\pm$0.18 MeV, derived from Coulomb mass shift
calculations~\citep{bro02}. It was found that the effective lifetime of $^{64}$Ge for a given temperature and proton density is mainly determined by the $S_p$
values of $^{65}$As and $^{66}$Se and the proton capture rate on $^{65}$As.

Recently, precise mass measurements of nuclei along the $rp$-process path have become available. The mass of $^{64}$Ge has been measured at the Canadian Penning Trap at Argonne National Laboratory~\citep{cla07} and the LEBIT Penning Trap facility at Michigan State University~\citep{sch07}. More recently, the mass of $^{65}$As has been measured at the HIRFL-CSR (Cooler-Storage Ring at the Heavy Ion Research Facility in Lanzhou)~\citep{xia02} using IMS (Isochronous Mass Spectrometry). The measurements can be combined to obtain an experimental proton separation energy for $^{65}$As of $S_p=-90\pm85$~keV~\citep{tu11}, where the uncertainty is dominated by the uncertainty in the $^{65}$As mass. The mass of $^{66}$Se is not known experimentally. The extrapolated value predicted by AME2012 results in $S_{\rm p}(^{66}$Se$)=1720\pm310$~keV. With the new mass of $^{65}$As, X-ray burst model calculations~\citep{tu11} suggested that $^{64}$Ge may not be a significant $rp$-process waiting point,  contrary to previous expectations~\citep{sch98,woo04,fis08,par09,jos10}. We revisit this question here using our new nuclear reaction rates.

Thermonuclear $^{64}$Ge($p$,$\gamma$) and $^{65}$As($p$,$\gamma$) reaction rates were first estimated by~\citet{wor94} based entirely on the properties of the mirror nuclei $^{65}$Ge and $^{66}$Ge, respectively. $S_p$ values of $^{65}$As and $^{66}$Se were estimated to be 0.169 MeV and 1.909 MeV, respectively. Later on, both rates have been calculated~\citep{rau00} with the statistical Hauser-Feshbach formalism (NON-SMOKER~\citep{rau98}) using the masses of $^{65}$As and $^{66}$Se predicted by the finite-range droplet (FRDM)~\citep{frdm95} and ETSFIQ~\citep{pea96} mass models. Recently, the statistical model calculations have been updated using new predictions for the $^{65}$As and $^{66}$Se proton separation energies (see JINA REACLIB\footnote{http://groups.nscl.msu.edu/jina/reaclib/db}~\citep{cyb10}). The predicted rates differ from one another by up to several orders of magnitude over typical XRB temperatures. Moreover, the reliability of statistical model calculations for these rates is questionable due to the low compound nucleus level densities, especially for $^{64}$Ge($p$,$\gamma$), but also for $^{65}$As($p$,$\gamma$).

In this work we refer to previously available rates using the nomenclature adopted in the JINA REACLIB database. The \emph{laur} rate refers to the rate estimated by~\citet{wor94}; the \emph{rath} rate was calculated by~\citet{rau00}. The \emph{rath, thra, rpsm} rates are the statistical-model calculations with FRDM, ETSFIQ, as well as~\citet{aud95} estimated masses, respectively. The recent \emph{ths8} rate is from \citet{cyb10}.

Here we determine new thermonuclear $^{64}$Ge($p$,$\gamma$)$^{65}$As and $^{65}$As($p$,$\gamma$)$^{66}$Se reaction rates using the updated $S_p$ values of $^{65}$As and $^{66}$Se together with new nuclear structure information from large-scale shell-model calculations. Using the new data we fully characterize the nuclear physics uncertainties that affect the $rp$-process through $^{64}$Ge and reexamine the question of the $^{64}$Ge waiting point.

\section{Reaction rate calculations}
The total thermonuclear proton capture reaction rate consists of the sum of resonant- and direct-capture (DC) on ground state and
thermally excited states in the target nucleus, weighted with their individual population factors~\citep{fow64,rol88}. It can be calculated by the following
equation:
\begin{eqnarray}
N_A\langle \sigma v \rangle = \sum_i(N_A\langle \sigma v \rangle_r^i+N_A\langle \sigma v \rangle_\mathrm{DC}^i)\frac{(2J_i+1)e^{-E_i/kT}}{\sum_n(2J_n+1)e^{-E_n/kT}} \nonumber \\
\label{eq1}
\end{eqnarray}
with the parameters defined by~\citet{sch05}.

\subsection{Resonant rates}
\label{sec:Resonant}
For isolated narrow resonances, the resonant reaction rate for capture on a nucleus in an initial state $i$, $N_A\langle \sigma v \rangle_r^i$, can be calculated
as a sum over all relevant compound nucleus states $j$ above the proton threshold~\citep{rol88,ili07}. It can be expressed by the following equation~\citep{sch05}:
\begin{eqnarray}
N_A\langle \sigma v \rangle_r^i = && 1.54 \times 10^{11} (\mu T_9)^{-3/2} \nonumber\\
                                && \times \sum_j \omega\gamma_{ij} \mathrm{exp} \left (-\frac{11.605E_{ij}}{T_9} \right) [\mathrm{cm^3s^{-1}mol^{-1}}], \nonumber \\
\label{eq2}
\end{eqnarray}
where the resonance energy in the center-of-mass system, $E_{ij}=E_j-S_p-E_i$, is calculated from the excitation energies of the initial $E_i$ and compound nucleus $E_j$ state. For the ground-state capture, the resonance energy is represented by $E_r^i=E_x^j-S_p$. $T_9$ is the temperature in Giga Kelvin (GK) and $\mu$ is the reduced mass of the entrance channel in atomic mass units ($\mu=A_T/(1+A_T)$, with $A_T$ the target mass number). In Eq.~\ref{eq2}, the resonance energy and strength are in units of MeV. The resonance strength $\omega\gamma$ is defined by 
\begin{eqnarray}
\omega\gamma_{ij}=\frac{2J_j+1}{2(2J_i+1)}\frac{\Gamma_p^{ij}\times\Gamma_\gamma^j}{\Gamma_\mathrm{total}^j}.
\label{eq3}
\end{eqnarray}
where $J_i$ is the target spin and $J_j$, $\Gamma_p^{ij}$ , $\Gamma_\gamma^j$, and $\Gamma_\mathrm{total}^j$ are spin, proton decay width, $\gamma$-decay width, and total width of the compound nucleus state $j$, respectively. The total width is given by $\Gamma_\mathrm{total}^j=\Gamma_\gamma^j+\sum_i\Gamma_p^{ij}$, because other decay channels are closed~\citep{aud12} in the excitation energy range considered in this work.

\begin{table*}[ht]
\begin{center}
\footnotesize
\caption{Properties of $^{65}$As for the ground-state capture utilized in the present $^{64}$Ge($p$,$\gamma$)$^{65}$As resonant rate calculation. The energy levels in the mirror nucleus $^{65}$Ge are listed in the 4$^\mathrm{th}$ column for comparison.}
\label{Table_65As}
\begin{tabular*}{\linewidth}{@{\hspace{2mm}\extracolsep{\fill}}lccccccccc@{\hspace{2mm}}}
\hline
\hline
&&&&&&&&&\\
$J^{\pi}_i$[a] & \multicolumn{3}{c}{$E_{x}$ [MeV] }                                                  &$E_r$ (MeV)[d]  & $nlj$   & $C^2S$    & $\Gamma_p$ [eV] & $\Gamma_\gamma$ [eV] & $\omega\gamma$ [eV]   \\
            & $E^\mathrm{exp}_x$[a] & $E^\mathrm{theo}_x$[b] & $E_x$($^{65}$Ge)[c]  &                        &         &           &                 &                      &                       \\
&&&&&&&&&\\
\hline
&&&&&&&&&\\
3/2$^{-}_1$ & 0.000  & 0.000  & 0.000  &  0.090     & 2$p_{3/2}$  & 0.196  & $1.19\times10^{-34}$                  & 0.00                  & 0.00                    \\
5/2$^{-}_1$[e] & 0.187(3)  & 0.103  & 0.111  &  0.277     & 1$f_{5/2}$  & 0.533  & $8.19\times10^{-17}$ & 5.11$\times$10$^{-7}$ & 2.46$\times$10$^{-16}$ \\
5/2$^{-}_2$ &        & 0.501  & 0.605  &  0.591     & 1$f_{5/2}$  & 0.010  & $3.76\times10^{-10}$                  & 4.73$\times$10$^{-5}$ & 1.13$\times$10$^{-9 }$  \\
5/2$^{-}_3$ &        & 0.863  &        &  0.953     & 1$f_{5/2}$  & 0.014  & $1.64\times10^{-6}$                   & 4.24$\times$10$^{-4}$ & 4.89$\times$10$^{-6 }$  \\
7/2$^{-}_1$ &        & 0.947  & 0.890  &  1.037     & 1$f_{7/2}$  & 0.013  & $1.28\times10^{-5}$                   & 2.43$\times$10$^{-4}$ & 4.88$\times$10$^{-5 }$  \\
7/2$^{-}_2$[f] &        & 1.070  & 1.155  &  1.160     & 1$f_{7/2}$  & 0.002  & $8.50\times10^{-6}$                   & 2.11$\times$10$^{-4}$ & 3.27$\times$10$^{-5 }$  \\
&&&&&&&&&\\
\hline
\end{tabular*}
\end{center}
\begin{minipage}[h]{\linewidth}
$[\textnormal{a}]$ measured by \citet{obe11};\\
$[\textnormal{b}]$ calculated by the present shell model;\\
$[\textnormal{c}]$ compiled by \citet{bro10};\\
$[\textnormal{d}]$ calculated by $E_r=E_x-S_p$ with $S_p=-0.09$ MeV~\citep{tu11};\\
$[\textnormal{e}]$ calculated $E_r$ and $\Gamma_p$ based on the experimental value of $E_x=0.187$~MeV for this state;\\
$[\textnormal{f}]$ negligible contribution to the rate for temperatures up to 2 GK.
\end{minipage}
\end{table*}

The proton width can be estimated by the following equation,
\begin{eqnarray}
\label{eq4}
\Gamma_p = \sum_{nlj} C^2S(nlj) \, \Gamma_{sp}(nlj) \, ,
\end{eqnarray}
where $C^2 S(nlj)$ denotes a proton-transfer spectroscopic factor, while $\Gamma_{sp}$ is a single-proton width for capture of a proton on an $(nlj)$ quantum
orbital. The $\Gamma_{sp}$ are obtained from proton scattering cross sections calculated with a Woods-Saxon potential~\citep{ric11,WSPOT}.
Alternatively, the proton partial widths may also be calculated by the following expression~\citep{wor94,her95},
\begin{eqnarray}
\Gamma_{p}=\frac{3\hbar^2}{\mu R^2}P_{\ell}(E)C^2S.
\label{eq5}
\end{eqnarray}
Here, $R=r_0\times(1+A_T)^{1/3}$ fm (with $r_0=1.25$~fm) is the nuclear channel radius. The Coulomb penetration factor $P_{\ell}$ is given by
\begin{eqnarray}
P_{\ell}(E)=\frac{kR}{F^2_{\ell}(E)+G^2_{\ell}(E)},
\label{eq6}
\end{eqnarray}
where $k=\sqrt{2\mu E}/\hbar$ is the wave number with energy $E$ in the center-of-mass (c.m.) system; $F_{\ell}$ and $G_{\ell}$ are the regular and irregular
Coulomb functions, respectively. The proton widths given by these two methods (i.e., by Eqs.~\ref{eq4} and~\ref{eq5}) agree well with each other, with a
maximum difference of about 35\%.

The key ingredients necessary to estimate the resonant $^{64}$Ge($p$,$\gamma$) and $^{65}$As($p$,$\gamma$) rates are energy levels in $^{65}$As and $^{66}$Se,
proton transfer spectroscopic factors, and proton and gamma-ray partial widths. For $^{65}$As, only a single level has been observed at
$E_x = 187(3)$~keV~\citep{obe11}. For $^{66}$Se, one level has been confirmed at $E_x = 929(7)$~keV, and indications for two other levels at
2064(3)~keV and 3520(4)~keV have been reported, with tentative assignments of (4$^+$) and (6$^+$), respectively~\citep{obe11,ruo13}. There are no more experimental data available for these two nuclei. In this work, we have calculated the energy levels, spectroscopic factors and gamma widths within the framework of a large-scale shell model, without truncation, using the shell-model code NuShellX@MSU~\citep{bro14}. The effective interaction GXPF1a~\citep{hon04,hon05} has been utilized for these two \textit{pf}-shell nuclei.

The $\gamma$ widths, $\Gamma_\gamma$, have been calculated from the electromagnetic reduced transition probabilities $B$($J_i \rightarrow J_f; L$), which carry the nuclear structure information of the resonant states and the final bound states~\citep{bru77}. The reduced transition rates are computed within the shell model. Most of the transitions in this work are of $M1$ and $E2$ types. The relations are~\citep{her95}:
\begin{eqnarray}
\Gamma_{E2}\mathrm{[eV]} = 8.13\times 10^{-7}E_\gamma^5 \mathrm{[MeV]}B(E2) \mathrm{[e^2fm^4]},
\label{eq51}
\end{eqnarray}
and
\begin{eqnarray}
\Gamma_{M1}\mathrm{[eV]} = 1.16\times 10^{-2}E_\gamma^3 \mathrm{[MeV]}B(M1) \mathrm{[\mu_N^2]}.
\label{eq52}
\end{eqnarray}
The $B(E2)$ values have been obtained from empirical effective charges, $e_p=1.5e$, $e_n=0.5e$, whereas the $B(M1)$ values have been obtained with a four-parameter set of empirical $g$-factors, i.e. $g^s_p=5.586$, $g^s_n=-3.826$ and $g^l_p=1$, $g^l_n=0$~\citep{hon04}.

In addition, we have also included proton resonant captures on the thermally excited target states. Since the first-excited state in $^{64}$Ge is quite high ($E_x=902$~keV), thermal excitation can be neglected for typical X-ray burst temperatures. For the $^{65}$As($p$,$\gamma$)$^{66}$Se rate, we included proton capture on the first four thermally excited states of $^{65}$As (i.e., on the 0.187, 0.501, 0.863 and 0.947 MeV states listed in Table~\ref{Table_65As}).
Capture on thermally excited states contributes at most 38\% to the total capture rate at 2 GK~\citep{lam16}.
The properties of $^{65}$As and $^{66}$Se for the ground-state captures are summarized in Table~\ref{Table_65As} and Table~\ref{Table_66Se}, respectively. In addition, the properties of $^{66}$Se for the first-excited-state capture (the major thermally excited-state contribution) are summarized in Table~\ref{Table_66Se02}.

\renewcommand{\arraystretch}{0.85}
\begin{table*}[ht]
\scriptsize
\caption{Properties of $^{66}$Se for the ground-state capture utilized in the present $^{65}$As($p$,$\gamma$)$^{66}$Se resonant rate calculation.}
\label{Table_66Se}
\begin{center}
\begin{tabular*}{\linewidth}{@{\hspace{2mm}\extracolsep{\fill}}lllllllllll@{\hspace{2mm}}}
\hline
\hline
&&&&&&&&&&\\
$J^{\pi}_i$[a] &\multicolumn{2}{c}{$E_{x}$ [MeV] }                 & $E_r$ [MeV][c]  & $C^2S_{7/2}$ & $C^2S_{3/2}$ & $C^2S_{5/2}$ &  $C^2S_{1/2}$         & $\Gamma_\gamma$ [eV]  & $\Gamma_p$ [eV]  & $\omega\gamma$ [eV] \\
  & $E_{x}^\mathrm{exp}$[b] & $E_{x}^\mathrm{theo}$[a]   &                 & $(l=3)$      & $(l=1)$      & $(l=3)$      &  $(l=1)$              &                       &                  &         \\
&&&&&&&&&&\\
\hline
&&&&&&&&&&\\
${0}^{+}_1$   & 0.000      & 0.000                &  ---           &         &  0.746  &         &         &       ---             &       ---              &      ---               \\
${2}^{+}_1$   & 0.929(7)   & 0.982                &  ---           & $0.037$ & $0.069$ & $0.119$ & $0.045$ & $2.765\times10^{-4 }$ &       ---              &      ---               \\
${0}^{+}_2$   &            & 1.130                &  ---           &         &  0.013  &         &         & $2.020\times10^{-8 }$ &       ---              &      ---               \\
${2}^{+}_2$   &            & 1.552                &  ---           & $0.002$ & $0.011$ & $0.071$ & $0.051$ & $1.402\times10^{-4 }$ &       ---              &      ---               \\
${0}^{+}_3$   &            & 1.575                &  ---           &         & $0.002$ &         &         & $2.056\times10^{-6 }$ &       ---              &      ---               \\
${2}^{+}_3$   &            & 1.952                & 0.232          & $0.001$ & $0.001$ & $0.180$ & $0.000$ & $4.040\times10^{-4 }$ & $3.079\times10^{-20 }$ & $1.924\times10^{-20 }$ \\
${0}^{+}_4$   &            & 1.989                & 0.269          &         & $0.002$ &         &         & $1.441\times10^{-8 }$ & $4.797\times10^{-18 }$ & $5.996\times10^{-19 }$ \\
${2}^{+}_4$   &            & 2.053                & 0.333[e]       & $0.002$ & $0.005$ & $0.027$ & $0.025$ & $2.218\times10^{-4 }$ & $3.517\times10^{-14 }$ & $2.198\times10^{-14 }$ \\
${4}^{+}_1$   & 2.064(3)[d]& 2.110                & 0.344          & $0.008$ &         & $0.004$ &         & $8.382\times10^{-4 }$ & $2.301\times10^{-16 }$ & $2.589\times10^{-16 }$ \\
${3}^{+}_1$   &            & 2.102                & 0.382          & $0.001$ & $0.000$ & $0.017$ &         & $3.835\times10^{-5 }$ & $1.057\times10^{-14 }$ & $9.249\times10^{-15 }$ \\
${2}^{+}_5$   &            & 2.277                & 0.557[e]       & $0.000$ & $0.004$ & $0.015$ & $0.001$ & $1.751\times10^{-4 }$ & $2.205\times10^{-9  }$ & $1.378\times10^{-9  }$ \\
${3}^{+}_2$   &            & 2.419                & 0.699          & $0.000$ & $0.004$ & $0.063$ &         & $2.361\times10^{-4 }$ & $1.781\times10^{-7  }$ & $1.557\times10^{-7  }$ \\
${1}^{+}_1$   &            & 2.474                & 0.754[e]       &         & $0.016$ & $0.000$ & $0.052$ & $3.579\times10^{-3 }$ & $1.152\times10^{-5  }$ & $4.306\times10^{-6  }$ \\
${4}^{+}_2$   &            & 2.498                & 0.778          & $0.001$ &         & $0.115$ &         & $2.653\times10^{-4 }$ & $1.888\times10^{-7  }$ & $2.122\times10^{-7  }$ \\
${2}^{+}_6$   &            & 2.556                & 0.836[e]       & $0.001$ & $0.021$ & $0.033$ & $0.032$ & $9.915\times10^{-4 }$ & $6.723\times10^{-5  }$ & $3.935\times10^{-5  }$ \\
${3}^{+}_3$   &            & 2.668                & 0.948          & $0.000$ & $0.065$ & $0.145$ &         & $1.550\times10^{-4 }$ & $7.969\times10^{-4  }$ & $1.136\times10^{-4  }$ \\
${4}^{+}_3$   &            & 2.740                & 1.020          & $0.002$ &         & $0.040$ &         & $8.404\times10^{-4 }$ & $8.287\times10^{-6  }$ & $9.232\times10^{-6  }$ \\
${1}^{+}_2$   &            & 2.781                & 1.061[e]       &         & $0.060$ & $0.351$ & $0.002$ & $5.338\times10^{-3 }$ & $4.483\times10^{-3  }$ & $9.137\times10^{-4  }$ \\
${2}^{+}_7$   &            & 2.865                & 1.145          & $0.000$ & $0.000$ & $0.000$ & $0.028$ & $1.175\times10^{-3 }$ & $5.315\times10^{-3  }$ & $6.014\times10^{-4  }$ \\
${1}^{+}_3$   &            & 2.867                & 1.147          &         & $0.065$ & $0.001$ & $0.222$ & $2.170\times10^{-3 }$ & $5.887\times10^{-2  }$ & $7.847\times10^{-4  }$ \\
${3}^{+}_4$   &            & 2.882                & 1.162          & $0.002$ & $0.003$ & $0.002$ &         & $1.209\times10^{-3 }$ & $9.439\times10^{-4  }$ & $4.639\times10^{-4  }$ \\
${0}^{+}_5$   &            & 2.888                & 1.168          &         & $0.629$ &         &         & $1.716\times10^{-4 }$ & $1.985\times10^{-1  }$ & $2.143\times10^{-5  }$ \\
${4}^{+}_4$   &            & 2.907                & 1.187          & $0.000$ &         & $0.005$ &         & $3.139\times10^{-4 }$ & $1.001\times10^{-5  }$ & $1.091\times10^{-5  }$ \\
${2}^{+}_8$   &            & 2.949                & 1.229          & $0.000$ & $0.003$ & $0.005$ & $0.016$ & $9.006\times10^{-4 }$ & $1.085\times10^{-2  }$ & $5.197\times10^{-4  }$ \\
${3}^{+}_5$   &            & 2.969                & 1.249          & $0.000$ & $0.006$ & $0.000$ &         & $7.804\times10^{-4 }$ & $4.817\times10^{-3  }$ & $5.877\times10^{-4  }$ \\
${4}^{+}_5$   &            & 2.998                & 1.278          & $0.000$ &         & $0.079$ &         & $1.392\times10^{-3 }$ & $4.711\times10^{-4  }$ & $3.960\times10^{-4  }$ \\
${4}^{+}_6$   &            & 3.114                & 1.394          & $0.000$ &         & $0.157$ &         & $1.374\times10^{-3 }$ & $3.262\times10^{-3  }$ & $1.088\times10^{-3  }$ \\
${2}^{+}_9$   &            & 3.217                & 1.497          & $0.000$ & $0.003$ & $0.021$ & $0.036$ & $1.534\times10^{-3 }$ & $3.221\times10^{-1  }$ & $9.539\times10^{-4  }$ \\
${0}^{+}_6$   &            & 3.231                & 1.511          &         & $0.002$ &         &         & $2.561\times10^{-3 }$ & $2.244\times10^{-2  }$ & $2.873\times10^{-4  }$ \\
${3}^{+}_6$   &            & 3.237                & 1.517          & $0.000$ & $0.000$ & $0.006$ &         & $1.586\times10^{-3 }$ & $5.258\times10^{-3  }$ & $1.066\times10^{-3  }$ \\
${1}^{+}_4$   &            & 3.274                & 1.554          &         & $0.005$ & $0.072$ & $0.010$ & $3.113\times10^{-3 }$ & $2.111\times10^{-1  }$ & $1.151\times10^{-3  }$ \\
${4}^{+}_7$   &            & 3.299                & 1.579          & $0.000$ &         & $0.008$ &         & $9.407\times10^{-4 }$ & $1.075\times10^{-3  }$ & $5.644\times10^{-4  }$ \\
${3}^{+}_7$   &            & 3.328                & 1.608          & $0.001$ & $0.004$ & $0.022$ &         & $1.139\times10^{-3 }$ & $8.762\times10^{-2  }$ & $9.842\times10^{-4  }$ \\
${2}^{+}_{10}$&            & 3.330                & 1.610          & $0.008$ & $0.083$ & $0.001$ & $0.012$ & $2.089\times10^{-3 }$ & $2.222\times10^{+0  }$ & $1.304\times10^{-3  }$ \\
${2}^{+}_{11}$&            & 3.357                & 1.637          & $0.001$ & $0.000$ & $0.027$ & $0.114$ & $1.348\times10^{-3 }$ & $2.761\times10^{+0  }$ & $8.423\times10^{-4  }$ \\
${5}^{+}_1 $  &            & 3.394                & 1.674          & $0.000$ &         &         &         & $1.355\times10^{-3 }$ & $1.460\times10^{-4  }$ & $1.812\times10^{-4  }$ \\
${4}^{+}_8 $  &            & 3.424                & 1.704          & $0.000$ &         & $0.018$ &         & $1.417\times10^{-3 }$ & $5.639\times10^{-3  }$ & $1.274\times10^{-3  }$ \\
${3}^{+}_8 $  &            & 3.487                & 1.767          & $0.000$ & $0.000$ & $0.056$ &         & $4.881\times10^{-3 }$ & $2.792\times10^{-2  }$ & $3.635\times10^{-3  }$ \\
${2}^{+}_{12}$&            & 3.499                & 1.779          & $0.006$ & $0.060$ & $0.001$ & $0.041$ & $4.335\times10^{-4 }$ & $7.440\times10^{+0  }$ & $2.709\times10^{-4  }$ \\
&&&&&&&&&&\\
\hline
\end{tabular*}
\end{center}
\begin{minipage}[h]{\linewidth}
$[\textnormal{a}]$calculated by the present shell model;\\
$[\textnormal{b}]$measured by~\citet{obe11} and~\citet{ruo13};\\
$[\textnormal{c}]$calculated by $E_r=E_x-S_p$ with $S_p=1.720$ MeV (AME2012);\\ 
$[\textnormal{d}]$calculated $E_r$ and $\Gamma_p$ based on the experimental value of $E_x=2.064$~MeV for this state;\\
$[\textnormal{e}]$resonances dominantly contributing to the rate within temperature region of 0.2--2 GK.
\end{minipage}
\end{table*}
\renewcommand{\arraystretch}{1.0}

\renewcommand{\arraystretch}{0.85}
\begin{table*}[ht]
\scriptsize
\caption{Properties of $^{66}$Se for the first-excited-state capture utilized in the present $^{65}$As$^{5/2_1^-}$($p$,$\gamma$)$^{66}$Se resonant rate.}
\label{Table_66Se02}
\begin{tabular*}{\linewidth}{@{\hspace{2mm}\extracolsep{\fill}}cllclllllll@{\hspace{2mm}}}
\hline
\hline
&&&&&&&&&&\\
$J^{\pi}_i$[a] &\multicolumn{2}{c}{$E_{x}$ [MeV] }                 & $E_r$ [MeV][c]  & $C^2S_{7/2}$ & $C^2S_{3/2}$ & $C^2S_{5/2}$ &  $C^2S_{1/2}$         & $\Gamma_\gamma$ [eV]  & $\Gamma_p$ [eV]  & $\omega\gamma$ [eV] \\
& $E_{x}^\mathrm{exp}$[b] & $E_{x}^\mathrm{theo}$[a]   &                 & $(l=3)$      & $(l=1)$      & $(l=3)$      &  $(l=1)$              &                       &                  &         \\
&&&&&&&&&&\\
\hline
&&&&&&&&&&\\
%
%
${2}^{+}_3$   &            & 1.952                & 0.045          &  0.002  &  0.035  &  0.000  &  0.013  & $4.040\times10^{-4 }$ & $4.401\times10^{-56}$ & $1.834\times10^{-56}$ \\
${0}^{+}_4$   &            & 1.989                & 0.082          &         &         &  0.244  &         & $1.441\times10^{-8 }$ & $2.398\times10^{-40}$ & $1.998\times10^{-41}$ \\
${2}^{+}_4$   &            & 2.053                & 0.146          &  0.000  &  0.019  &  0.088  &  0.098  & $2.218\times10^{-4 }$ & $4.499\times10^{-26}$ & $1.875\times10^{-26}$ \\
${4}^{+}_1$   & 2.064(3)[d]& 2.110                & 0.157          &  0.001  &  0.016  &  0.002  &         & $8.382\times10^{-4 }$ & $1.261\times10^{-25}$ & $9.458\times10^{-26}$ \\
${3}^{+}_1$   &            & 2.102                & 0.195          &  0.001  &  0.011  &  0.000  &  0.001  & $3.835\times10^{-5 }$ & $4.039\times10^{-22}$ & $2.356\times10^{-22}$ \\
${2}^{+}_5$   &            & 2.277                & 0.370          &  0.001  &  0.038  &  0.020  &  0.184  & $1.751\times10^{-4 }$ & $4.549\times10^{-12}$ & $1.895\times10^{-12}$ \\
${3}^{+}_2$   &            & 2.419[e]             & 0.512          &  0.001  &  0.083  &  0.004  &  0.226  & $2.361\times10^{-4 }$ & $1.862\times10^{-8 }$ & $1.086\times10^{-8 }$ \\
${1}^{+}_1$   &            & 2.474                & 0.567          &  0.000  &  0.000  &  0.002  &         & $3.579\times10^{-3 }$ & $5.149\times10^{-11}$ & $1.287\times10^{-11}$ \\
${4}^{+}_2$   &            & 2.498                & 0.591          &  0.000  &  0.043  &  0.208  &         & $2.653\times10^{-4 }$ & $6.133\times10^{-8 }$ & $4.599\times10^{-8 }$ \\
${2}^{+}_6$   &            & 2.556                & 0.649          &  0.001  &  0.026  &  0.191  &  0.008  & $9.915\times10^{-4 }$ & $3.292\times10^{-7 }$ & $1.371\times10^{-7 }$ \\
${3}^{+}_3$   &            & 2.668[e]             & 0.761          &  0.004  &  0.107  &  0.000  &  0.002  & $1.550\times10^{-4 }$ & $2.983\times10^{-5 }$ & $1.459\times10^{-5 }$ \\
${4}^{+}_3$   &            & 2.740                & 0.833          &  0.002  &  0.011  &  0.001  &         & $8.404\times10^{-4 }$ & $1.353\times10^{-5 }$ & $9.987\times10^{-6 }$ \\
${1}^{+}_2$   &            & 2.781                & 0.874          &  0.001  &  0.032  &  0.002  &         & $5.338\times10^{-3 }$ & $9.809\times10^{-5 }$ & $2.408\times10^{-5 }$ \\
${2}^{+}_7$   &            & 2.865[e]             & 0.958          &  0.000  &  0.002  &  0.324  &  0.054  & $1.175\times10^{-3 }$ & $6.788\times10^{-4 }$ & $1.793\times10^{-4 }$ \\
${1}^{+}_3$   &            & 2.867                & 0.960          &  0.000  &  0.000  &  0.004  &         & $2.170\times10^{-3 }$ & $2.502\times10^{-7 }$ & $6.254\times10^{-8 }$ \\
${3}^{+}_4$   &            & 2.882[e]             & 0.975          &  0.001  &  0.008  &  0.001  &  0.035  & $1.209\times10^{-3 }$ & $7.087\times10^{-4 }$ & $2.607\times10^{-4 }$ \\
${0}^{+}_5$   &            & 2.888                & 0.981          &         &         &  0.089  &         & $1.716\times10^{-4 }$ & $8.636\times10^{-6 }$ & $6.852\times10^{-7 }$ \\
${4}^{+}_4$   &            & 2.907                & 1.000          &  0.000  &  0.000  &  0.518  &         & $3.139\times10^{-4 }$ & $6.881\times10^{-5 }$ & $4.233\times10^{-5 }$ \\
${2}^{+}_8$   &            & 2.949                & 1.042          &  0.000  &  0.006  &  0.012  &  0.007  & $9.006\times10^{-4 }$ & $6.740\times10^{-4 }$ & $1.606\times10^{-4 }$ \\
${3}^{+}_5$   &            & 2.969[e]             & 1.062          &  0.002  &  0.003  &  0.001  &  0.025  & $7.804\times10^{-4 }$ & $1.754\times10^{-3 }$ & $3.151\times10^{-4 }$ \\
${4}^{+}_5$   &            & 2.998[e]             & 1.091          &  0.003  &  0.025  &  0.000  &         & $1.392\times10^{-3 }$ & $2.795\times10^{-3 }$ & $6.970\times10^{-4 }$ \\
${4}^{+}_6$   &            & 3.114[e]             & 1.207          &  0.001  &  0.061  &  0.103  &         & $1.374\times10^{-3 }$ & $3.138\times10^{-2 }$ & $9.872\times10^{-4 }$ \\
${2}^{+}_9$   &            & 3.217                & 1.310          &  0.000  &  0.019  &  0.016  &  0.001  & $1.534\times10^{-3 }$ & $3.314\times10^{-2 }$ & $6.107\times10^{-4 }$ \\
${0}^{+}_6$   &            & 3.231                & 1.324          &         &         &  0.025  &         & $2.561\times10^{-3 }$ & $2.545\times10^{-4 }$ & $1.929\times10^{-5 }$ \\
${3}^{+}_6$   &            & 3.237                & 1.330          &  0.001  &  0.007  &  0.000  &  0.053  & $1.586\times10^{-3 }$ & $1.026\times10^{-1 }$ & $9.112\times10^{-4 }$ \\
${1}^{+}_4$   &            & 3.274                & 1.367          &  0.002  &  0.002  &  0.002  &         & $3.113\times10^{-3 }$ & $4.840\times10^{-3 }$ & $4.737\times10^{-4 }$ \\
${4}^{+}_7$   &            & 3.299                & 1.392          &  0.000  &  0.003  &  0.026  &         & $9.407\times10^{-4 }$ & $1.039\times10^{-2 }$ & $6.470\times10^{-4 }$ \\
${3}^{+}_7$   &            & 3.328                & 1.421          &  0.001  &  0.073  &  0.003  &  0.019  & $1.139\times10^{-3 }$ & $4.284\times10^{-1 }$ & $6.629\times10^{-4 }$ \\
${2}^{+}_{10}$&            & 3.330                & 1.423          &  0.000  &  0.015  &  0.036  &  0.001  & $2.089\times10^{-3 }$ & $7.650\times10^{-2 }$ & $8.473\times10^{-4 }$ \\
${2}^{+}_{11}$&            & 3.357                & 1.450          &  0.000  &  0.020  &  0.001  &  0.006  & $1.348\times10^{-3 }$ & $1.496\times10^{-1 }$ & $5.568\times10^{-4 }$ \\
${5}^{+}_1 $  &            & 3.394                & 1.487          &  0.002  &         &  0.000  &         & $1.355\times10^{-3 }$ & $3.592\times10^{+3 }$ & $1.242\times10^{-3 }$ \\
${4}^{+}_8 $  &            & 3.424                & 1.517          &  0.000  &  0.013  &  0.145  &         & $1.417\times10^{-3 }$ & $1.591\times10^{-1 }$ & $1.054\times10^{-3 }$ \\
${3}^{+}_8 $  &            & 3.487                & 1.580          &  0.000  &  0.051  &  0.000  &  0.001  & $4.881\times10^{-3 }$ & $9.800\times10^{-1 }$ & $2.833\times10^{-3 }$ \\
${2}^{+}_{12}$&            & 3.499                & 1.592          &  0.001  &  0.006  &  0.041  &  0.008  & $4.335\times10^{-4 }$ & $2.659\times10^{-1 }$ & $1.803\times10^{-4 }$ \\
&&&&&&&&&&\\
\hline
\end{tabular*}
\begin{minipage}[h]{\linewidth}
$[\textnormal{a}]$calculated by the present shell model;\\
$[\textnormal{b}]$measured by~\citet{obe11} and~\citet{ruo13};\\
$[\textnormal{c}]$calculated by $E_r=E_x-S_p-0.187$ in units of MeV, with $S_p=1.720$ MeV (AME2012);\\
$[\textnormal{d}]$calculated $E_r$ and $\Gamma_p$ based on the experimental value of $E_x=2.064$~MeV for this state;\\
$[\textnormal{e}]$resonances dominantly contributing to the rate within temperature region of 0.2--2 GK.\\
\vspace{1.0cm}
\end{minipage}
\end{table*}
\renewcommand{\arraystretch}{1.0}

Peak temperatures in recent hydrodynamic XRB models have approached 1.5--2~GK~\citep{woo04,jos10}. At such temperatures, resonant rates for the
$^{64}$Ge($p$,$\gamma$) and $^{65}$As($p$,$\gamma$) reactions are expected to be dominated by levels with $E_r \leq2.5$~MeV (i.e., Gamow energy~\citep{rol88}).
This means that excitation energy regions of up to $E_x\leq$~2.5 MeV for $^{65}$As, and up to $E_x\leq$~4.2 MeV for $^{66}$Se should be considered in the
resonant rate calculations for $^{64}$Ge($p$,$\gamma$) and $^{65}$As($p$,$\gamma$), respectively. In the present shell-model calculations, the maximum excitation
energies considered for $^{65}$As and $^{66}$Se are 1.07 MeV and 3.50 MeV, respectively (see Tables~\ref{Table_65As} and~\ref{Table_66Se}).
For $^{64}$Ge($p$,$\gamma$) only resonances up to $E_r$=1.035 MeV contribute significantly to the reaction rate up to 2 GK; for $^{65}$As($p$,$\gamma$) only
five resonances (at $E_r$=0.333, 0.557, 0.754, 0.836 and 1.061 MeV) dominate the total resonant rate within the temperature region of 0.2--2 GK. Those 21
resonances above $E_r$=1.061 MeV make only negligible contributions to the total reaction rate up to 2 GK. Therefore, the contributions from the levels
presented in Tables~\ref{Table_65As} and~\ref{Table_66Se} should be adequate to account for these two resonant rates at XRB temperatures.

\subsection{Direct-capture rates}
\label{sec:DC}
The nonresonant direct-capture (DC) rate for proton capture can be estimated by the following expression~\citep{ang99,sch05},
\begin{eqnarray}
N_A\langle \sigma v \rangle_\mathrm{DC}^i= && 7.83 \times 10^{9} \left( \frac{Z_T}{\mu T_9^2} \right)^{1/3} S_\mathrm{DC}^i(E_0)  \nonumber \\
                                           && \times \mathrm{exp} \left[ -4.249 \left (\frac{Z_T^2 \mu}{T_9} \right)^{1/3} \right] [\mathrm{cm^3s^{-1}mol^{-1}}], \nonumber \\
\label{eq7}
\end{eqnarray}
with $Z_T$ being the atomic number of either $^{64}$Ge or $^{65}$As. The effective astrophysical $S$-factor at the Gamow energy $E_0$, i.e.,
$S_\mathrm{DC}^i(E_0)$, can be expressed by~\citep{fow64,rol88},
\begin{eqnarray}
S_\mathrm{DC}^i(E_0)=S^i(0)\left( 1+\frac{5}{12\tau} \right),
\label{eq8}
\end{eqnarray}
where $S^i$(0) is the $S$-factor at zero energy, and the dimensionless parameter $\tau$ is given numerically by $\tau=4.2487(Z_T^2 \mu/T_9)^{-1/3}$ for the
proton capture.

In this work, we have calculated the DC $S$-factors with the RADCAP code~\citep{ber03}. The Woods-Saxon nuclear potential (central + spin orbit) and a
Coulomb potential of uniform-charge distribution were utilized in the calculation. The nuclear central potential $V_0$ was determined by matching the bound-state
energies. The spectroscopic factors were taken from the shell model calculation and are listed in Table~\ref{Table_65As} and Table~\ref{Table_66Se}.
The optical-potential parameters~\citep{hua10} are $R_0=R_\mathrm{s.o.}=R_C=1.25\times(1+A_T)^{1/3}$~fm, $a_0=a_\mathrm{s.o.}=0.65$~fm, with a spin-orbit
potential depth of $V_\mathrm{s.o.}=-10$ MeV. Here, $R_0$, $R_\mathrm{s.o.}$, and $R_C$ are the radii of central potential, the spin-orbit potential and the
Coulomb potential, respectively; $a_0$ and $a_\mathrm{s.o.}$ are the corresponding diffuseness parameters in the central and spin-orbit potentials, respectively.

For the $^{65}$As($p$,$\gamma$)$^{66}$Se reaction, $S$(0) values for DC captures into the ground state and the first-excited state ($E_x$=929 keV) in
$^{66}$Se are calculated to be 8.3 and 3.5 MeV$\cdot$b, respectively. The total DC rate for this reaction is only about 0.1\% that of the resonant one at 0.05 GK.
For the $^{64}$Ge($p$,$\gamma$)$^{65}$As reaction, we find a DC $S$(0) value for this reaction of about 35 MeV$\cdot$b. The DC contribution is only about 0.3\%
even at the lowest temperature of 0.06 GK. Even when considering estimated upper limits to the DC contribution~\citep{hjj14}, the resonant contributions
still dominate the total rates above 0.06 GK and 0.05 GK for $^{64}$Ge($p$,$\gamma$)$^{65}$As and $^{65}$As($p$,$\gamma$)$^{66}$Se reactions, respectively.
The probabilities of populating the first-excited states in $^{64}$Ge ($E_x$=902 keV) and $^{65}$As ($E_x$=187 keV) relative to the ground
states at temperatures below 0.1 GK are extremely small, and hence contributions of the direct-capture from these excited states can be neglected.


\section{Results}
\label{sec:Total}
The resulting total thermonuclear $^{64}$Ge($p$,$\gamma$)$^{65}$As and $^{65}$As($p$,$\gamma$)$^{66}$Se rates are listed in Table~\ref{tab:rates} as functions of temperature. The present (\emph{Present}, hereafter) rates can be parameterized by the standard format of~\citet{rau00}.
For $^{64}$Ge($p$,$\gamma$)$^{65}$As, we find:
\begin{widetext}
\begin{eqnarray*}
N_A\langle\sigma v\rangle &=& \mathrm{exp}(-78.204-\frac{13.819}{T_9}+\frac{12.211}{T_9^{1/3}}+81.566T_9^{1/3}-13.138T_9+1.717T_9^{5/3}-16.149\ln{T_9}) \nonumber \\
                          &+& \mathrm{exp}(-93.260-\frac{10.059}{T_9}+\frac{15.189}{T_9^{1/3}}+61.887T_9^{1/3}+21.717T_9-8.625T_9^{5/3}-22.943\ln{T_9}) \nonumber \\
                          &+& \mathrm{exp}(-75.104-\frac{3.788}{T_9}+\frac{19.347}{T_9^{1/3}}+52.700T_9^{1/3}-32.227T_9+14.766T_9^{5/3}+1.270\ln{T_9}) \, ,
\label{eq9}
\end{eqnarray*}
\end{widetext}
with a fitting error of less than 0.3\% at 0.1--2 GK; for $^{65}$As($p$,$\gamma$)$^{66}$Se, we find:
\begin{widetext}
\begin{eqnarray*}
N_A\langle\sigma v\rangle &=& \mathrm{exp}(-111.177-\frac{2.639}{T_9}+\frac{50.997}{T_9^{1/3}}+106.669T_9^{1/3}-64.623T_9+13.521T_9^{5/3}+31.256\ln{T_9}) \nonumber \\
                          &+& \mathrm{exp}(-124.702-\frac{12.436}{T_9}+\frac{52.765}{T_9^{1/3}}+89.593T_9^{1/3}-12.219T_9+0.456T_9^{5/3}-2.886\ln{T_9}) \nonumber \\
                          &+& \mathrm{exp}(-116.814-\frac{5.202}{T_9}+\frac{63.424}{T_9^{1/3}}+48.281T_9^{1/3}+83.320T_9-188.849T_9^{5/3}+21.362\ln{T_9}) \, ,
\label{eq10}
\end{eqnarray*}
\end{widetext}
with a fitting error of less than 0.4\% at 0.1--2 GK. We emphasize that the above fits are only valid with the stated error over the temperature range of 0.1--2 GK.
Above 2 GK, one may, for example, match our rates to statistical model calculations (see e.g., NACRE by~\citet{ang99}).

\begin{table*}
\footnotesize
\caption{\label{Table_rate} Thermonuclear $^{64}$Ge($p$,$\gamma$)$^{65}$As and $^{65}$As($p$,$\gamma$)$^{66}$Se rates in units of \MakeLowercase{cm$^{3}$s$^{-1}$mol$^{-1}$}.}
\label{tab:rates}
\begin{tabular*}{\linewidth}{@{\hspace{2mm}\extracolsep{\fill}}ccccccc@{\hspace{2mm}}}
\hline
\hline
&&&&&&\\
&\multicolumn{3}{c}{$^{64}$Ge($p$,$\gamma$)$^{65}$As } & \multicolumn{3}{c}{$^{65}$As($p$,$\gamma$)$^{66}$Se } \\
$T_{9}$ & $N_{A}\langle\sigma v\rangle$ & lower & upper      & $N_{A}\langle\sigma v\rangle$ & lower & upper      \\
&&&&&&\\
\hline
&&&&&&\\
0.2  &   $4.79\times10^{-17}$ &   $3.99\times10^{-20}$ & $1.23\times10^{-15}$  & $1.86\times10^{-16}$ &    $9.97\times10^{-18}$ &   $1.45\times10^{-15}$  \\
0.3  &   $1.33\times10^{-13}$ &   $8.44\times10^{-15}$ & $7.66\times10^{-13}$  & $2.21\times10^{-12}$ &    $1.79\times10^{-12}$ &   $1.92\times10^{-11}$  \\
0.4  &   $3.08\times10^{-11}$ &   $9.92\times10^{-13}$ & $6.56\times10^{-11}$  & $1.96\times10^{-9}$  &    $8.15\times10^{-10}$ &   $8.26\times10^{-9 }$   \\
0.5  &   $1.89\times10^{-9}$  &   $1.98\times10^{-10}$ & $4.30\times10^{-9 }$  & $1.51\times10^{-7}$  &    $3.03\times10^{-8 }$ &   $4.65\times10^{-7 }$   \\
0.6  &   $5.47\times10^{-8}$  &   $1.42\times10^{-8 }$ & $8.93\times10^{-8 }$  & $3.08\times10^{-6}$  &    $3.31\times10^{-7 }$ &   $7.55\times10^{-6 }$   \\
0.7  &   $6.86\times10^{-7}$  &   $2.61\times10^{-7 }$ & $9.38\times10^{-7 }$  & $2.93\times10^{-5}$  &    $1.73\times10^{-6 }$ &   $5.91\times10^{-5 }$   \\
0.8  &   $4.62\times10^{-6}$  &   $2.15\times10^{-6 }$ & $6.52\times10^{-6 }$  & $1.69\times10^{-4}$  &    $6.47\times10^{-6 }$ &   $2.94\times10^{-4 }$   \\
0.9  &   $2.02\times10^{-5}$  &   $8.05\times10^{-6 }$ & $3.52\times10^{-5 }$  & $6.86\times10^{-4}$  &    $2.18\times10^{-5 }$ &   $1.08\times10^{-3 }$   \\
1.0  &   $6.53\times10^{-5}$  &   $2.17\times10^{-5 }$ & $1.39\times10^{-4 }$  & $2.15\times10^{-3}$  &    $7.04\times10^{-5 }$ &   $3.19\times10^{-3 }$   \\
1.1  &   $1.69\times10^{-4}$  &   $4.69\times10^{-5 }$ & $4.30\times10^{-4 }$  & $5.55\times10^{-3}$  &    $2.12\times10^{-4 }$ &   $7.96\times10^{-3 }$   \\
1.2  &   $3.69\times10^{-4}$  &   $8.83\times10^{-5 }$ & $1.09\times10^{-3 }$  & $1.23\times10^{-2}$  &    $4.67\times10^{-4 }$ &   $1.74\times10^{-2 }$   \\
1.3  &   $7.10\times10^{-4}$  &   $1.50\times10^{-4 }$ & $2.38\times10^{-3 }$  & $2.42\times10^{-2}$  &    $9.89\times10^{-4 }$ &   $3.41\times10^{-2 }$   \\
1.4  &   $1.24\times10^{-3}$  &   $2.34\times10^{-4 }$ & $4.62\times10^{-3 }$  & $4.33\times10^{-2}$  &    $1.90\times10^{-3 }$ &   $6.11\times10^{-2 }$   \\
1.5  &   $1.98\times10^{-3}$  &   $3.42\times10^{-4 }$ & $8.16\times10^{-3 }$  & $7.18\times10^{-2}$  &    $3.33\times10^{-3 }$ &   $1.02\times10^{-1 }$   \\
1.6  &   $2.98\times10^{-3}$  &   $4.73\times10^{-4 }$ & $1.33\times10^{-2 }$  & $1.12\times10^{-1}$  &    $5.69\times10^{-3 }$ &   $1.59\times10^{-1 }$   \\
1.7  &   $4.25\times10^{-3}$  &   $6.28\times10^{-4 }$ & $2.04\times10^{-2 }$  & $1.65\times10^{-1}$  &    $9.38\times10^{-3 }$ &   $2.34\times10^{-1 }$   \\
1.8  &   $5.78\times10^{-3}$  &   $8.02\times10^{-4 }$ & $2.97\times10^{-2 }$  & $2.34\times10^{-1}$  &    $1.49\times10^{-2 }$ &   $3.31\times10^{-1 }$   \\
1.9  &   $7.58\times10^{-3}$  &   $9.93\times10^{-4 }$ & $4.13\times10^{-2 }$  & $3.18\times10^{-1}$  &    $2.22\times10^{-2 }$ &   $4.50\times10^{-1 }$   \\
2.0  &   $9.63\times10^{-3}$  &   $1.20\times10^{-3 }$ & $5.52\times10^{-2 }$  & $4.19\times10^{-1}$  &    $3.21\times10^{-2 }$ &   $5.93\times10^{-1 }$   \\
&&&&&&\\
\hline
\end{tabular*}
\end{table*}

\begin{figure}[b]
\begin{center}
\includegraphics[width=8.8cm]{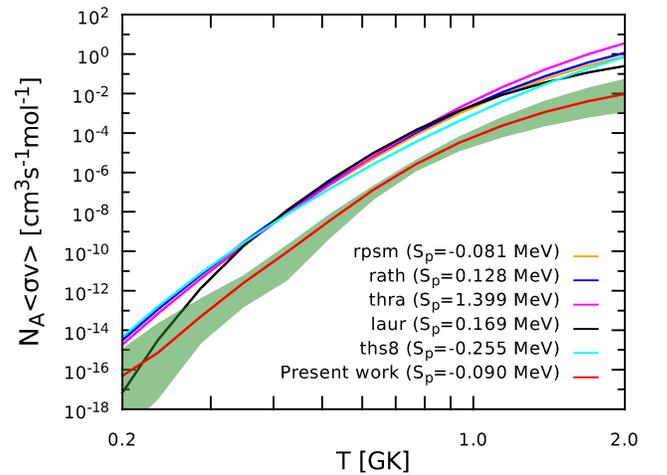}
\vspace{-4mm}
\caption{\label{Fig1_64Gep} Reaction rates of the $^{64}$Ge($p$,$\gamma$)$^{65}$As reaction (in units of cm$^3$ mol$^{-1}$ s$^{-1}$). The \emph{Present} rate
(red line) together with the upper and lower limits deduced from uncertainties are shown by a (green) colored band. Other available rates from JINA
REACLIB~\citep{cyb10} are shown for comparison. See details in the text and Table~\ref{tab:rates}. (A color version of this figure is available in the online
journal.)}
\end{center}
\end{figure}

\begin{figure}[b]
\begin{center}
\includegraphics[width=8.8cm]{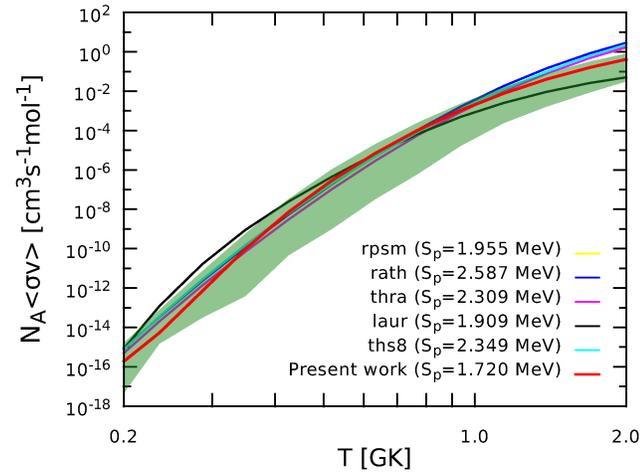}
\vspace{-4mm}
\caption{\label{Fig2_65Asp} As Fig.~\ref{Fig1_64Gep}, but for the $^{65}$As($p$,$\gamma$)$^{66}$Se reaction. Note the \emph{rpsm} rate is quite close to the \emph{rath}, \emph{ths8} and \emph{thra} rates at different temperature regions~\citep{cyb10}, and hence the corresponding line is not clearly visible. See details in the text and Table~\ref{tab:rates}. (A color version of this figure is available in the online journal.)}
\end{center}
\end{figure}

Figure~\ref{Fig1_64Gep} shows the comparison of the \emph{Present} $^{64}$Ge($p$,$\gamma$)$^{65}$As rate with others compiled in JINA REACLIB:
\emph{rpsm}, \emph{rath}, \emph{thra}, \emph{laur}, and \emph{ths8}. Note that only the \emph{rpsm} rate uses an $S_p$($^{65}$As) value that is within 1$\sigma$
of the recently determined experimental value. The \emph{Present} rate differs significantly from others in the temperature region of interest in XRBs.
The disagreement, in particular with the \emph{rpsm} rate, demonstrates that the statistical-model is not applicable for this reaction
owing to the low density of excited states in $^{65}$As.

Similarly, the comparison of the \emph{Present} $^{65}$As($p$,$\gamma$)$^{66}$Se rate with other rates available in the JINA REACLIB:
\emph{rpsm}, \emph{rath}, \emph{thra}, \emph{laur}, and \emph{ths8}, is presented in Fig.~\ref{Fig2_65Asp}. Only the \emph{laur} and \emph{rpsm} rates use
$S_p$($^{66}$Se) values that are within 1$\sigma$ of the currently accepted value. Although the \emph{Present} rate differs significantly from the others,
especially at lower and higher temperature regions, it is consistent with all others within the remaining large uncertainties. At $T>$1 GK, the \emph{laur} rate
is the lowest rate simply because only three excited states were considered by~\citet{wor94}. It should be noted that the shell model calculation provides the
first reliable estimate of the uncertainty of the $^{65}$As($p$,$\gamma$) reaction rate, especially as the Hauser-Feshbach rates may suffer from unknown
systematic errors due to the limited applicability of the statistical model near the proton drip line.

Uncertainties for the \emph{Present} $^{64}$Ge($p$,$\gamma$)$^{65}$As and $^{65}$As($p$,$\gamma$)$^{66}$Se rates were estimated by considering the
uncertainties in the $S_p$ values ($\pm$85 keV for $^{65}$As and $\pm$310 keV for $^{66}$Se) and estimated uncertainties in the calculated level energies
($\pm$168 keV for both $^{65}$As and $^{66}$Se~\citep{hon02}\footnote{The $^{66}$Zn case is studied with the present model space and interaction, and
an $rms$ deviation between the experimental and calculated level energies is found to be about 140 keV~\citep{lam16}.}). These were added in quadrature to give
uncertainties of $\pm$188 keV and $\pm$353 keV for the resonance energies $E_r$ of $^{64}$Ge($p$,$\gamma$)$^{65}$As and $^{65}$As($p$,$\gamma$)$^{66}$Se,
respectively. Note that for the two known levels, i.e., $E_x$=187 keV in $^{65}$As and $E_x$=2064 keV in $^{66}$Se, an experimental excitation energy uncertainty
of $\pm$3 keV is used instead. For $^{64}$Ge($p$,$\gamma$)$^{65}$As, all resonance strengths $\omega\gamma$ are proportional to $\Gamma_p$ since
$\Gamma_p$$\ll$$\Gamma_\gamma$ (see Table~\ref{Table_65As} and Eq.~\ref{eq3}). The uncertainties in $\Gamma_p$ (or $\omega\gamma$) owing to the uncertainties
in $E_r$ are calculated based on the energy dependence expressed in Eq.~\ref{eq5}. In the case of $^{65}$As($p$,$\gamma$)$^{66}$Se, only five resonances (at
$E_r$=0.333, 0.557, 0.754, 0.836 and 1.061 MeV dominate the resonant rate within the temperature region of 0.2--2 GK. Here, the resonant strengths
$\omega\gamma$ are proportional to $\Gamma_p$ for the first four resonances, while the strength of the last resonance at 1.061 MeV (dominating over
$\sim$0.9--2 GK) depends on both, $\Gamma_p$ and $\Gamma_\gamma$. Uncertainties in $\Gamma_\gamma$ can be neglected because of the much larger rate
uncertainties caused by $E_r$ and $\Gamma_p$. The uncertainties for all resonances listed in Table~\ref{Table_66Se} are considered in the present calculations.

\begin{figure}[t]
\begin{center}
\includegraphics[scale=0.65]{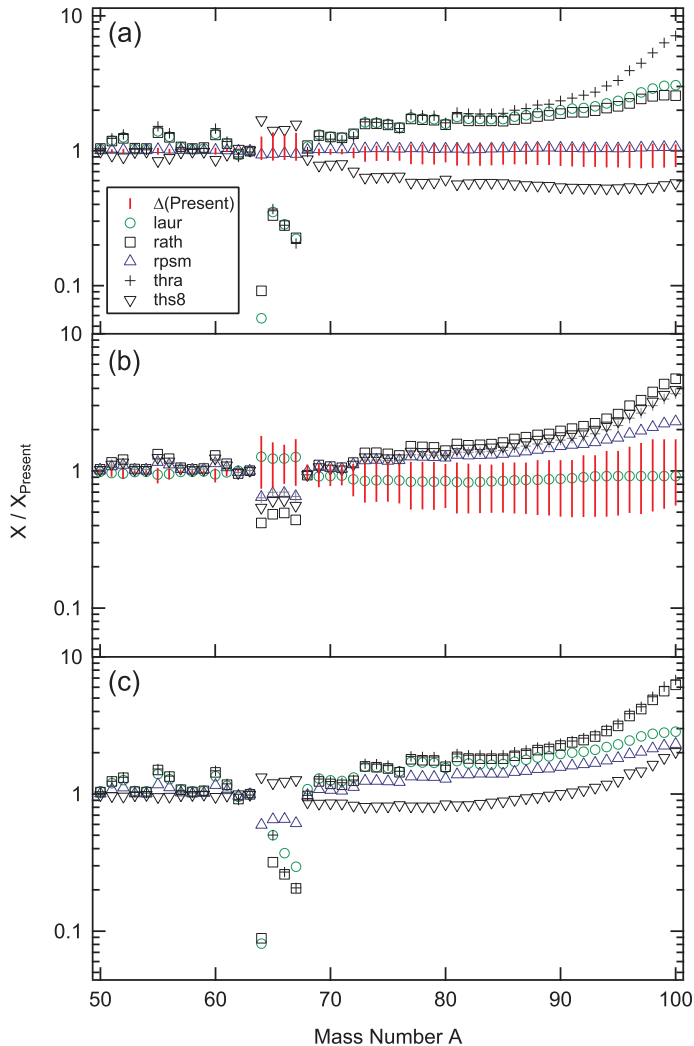}
\caption{Final abundances, as mass fractions $X$, following one-zone XRB calculations using the K04 thermodynamic history~\citep{par08, par09}. Results using
rates determined in the \emph{Present} work and in JINA REACLIB: \emph{rpsm}, \emph{rath}, \emph{thra}, \emph{laur}, and \emph{ths8} are indicated. Panels (a)
and (b) show the effects of using either different $^{64}$Ge($p$,$\gamma$) or different $^{65}$As($p$,$\gamma$) rates, respectively, while panel (c) shows the
effect of using different $^{64}$Ge($p$,$\gamma$) and $^{65}$As($p$,$\gamma$) rates together. The impact of the uncertainties in the \emph{Present}
$^{64}$Ge($p$,$\gamma$) and $^{65}$As($p$,$\gamma$) rates (see Figs.~\ref{Fig1_64Gep} and~\ref{Fig2_65Asp}) is indicated as \emph{$\Delta$(Present)} in
panels (a) and (b). (A color version of this figure is available in the online journal.)}
\label{fig3_Abund}
\end{center}
\end{figure}

\begin{figure*}[t]
\begin{center}
\includegraphics[scale=0.7]{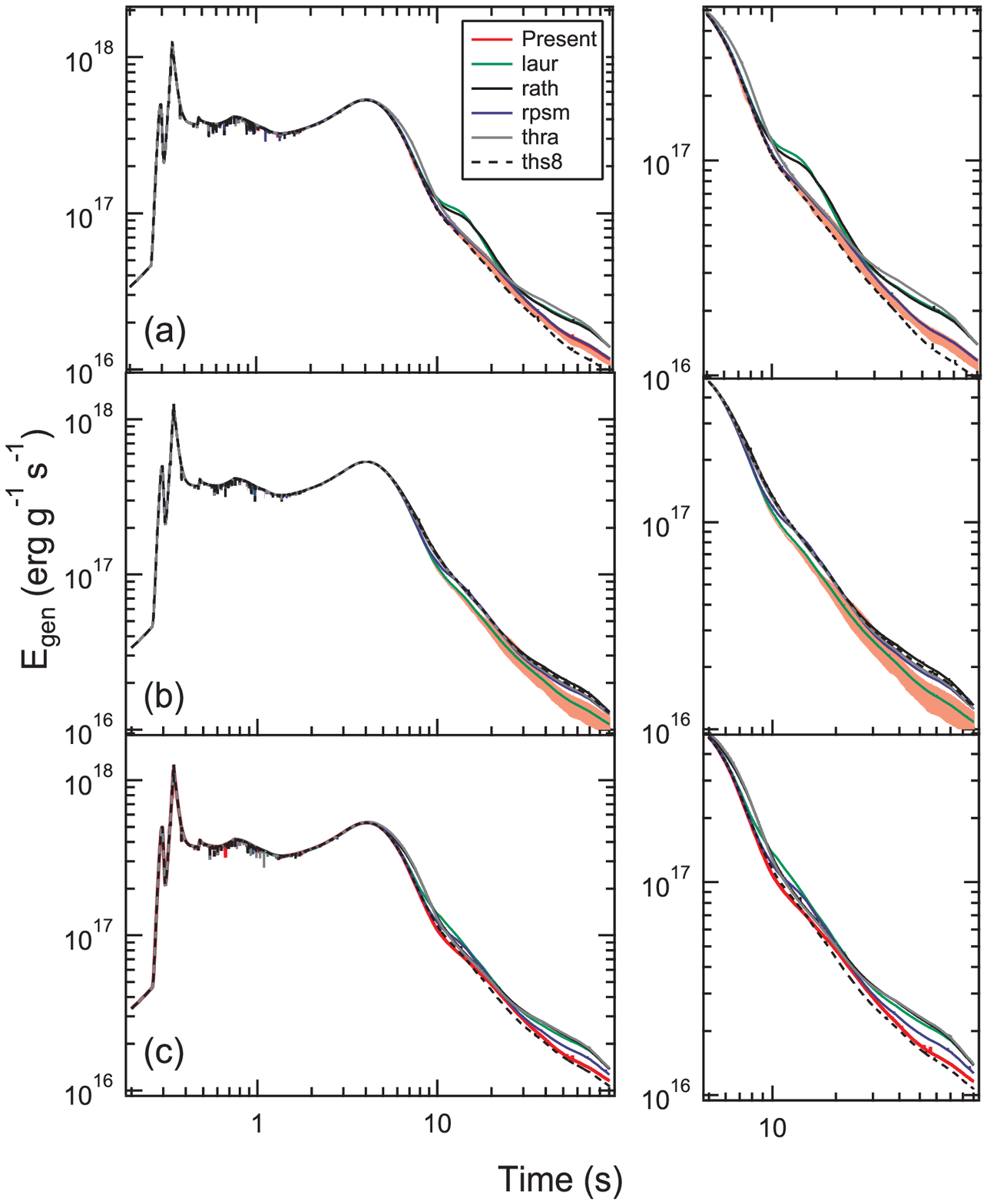}
\caption{As Fig.~\ref{fig3_Abund}, but for nuclear energy generation rates $E_\mathrm{gen}$ during the burst in the K04 XRB model. Panels (a) and (b) show the
effects of using either different $^{64}$Ge($p$,$\gamma$) or different $^{65}$As($p$,$\gamma$) rates, respectively, while panel (c) shows the effect of using
different $^{64}$Ge($p$,$\gamma$) and $^{65}$As($p$,$\gamma$) rates together. The impact of the uncertainties in the \emph{Present} $^{64}$Ge($p$,$\gamma$) and
$^{65}$As($p$,$\gamma$) rates (see Figs.~\ref{Fig1_64Gep} and~\ref{Fig2_65Asp}) is indicated in panels (a) and (b). Panels to the right show expanded views of
the panels to the left. (A color version of this figure is available in the online journal.)}
\label{fig4_Egen}
\end{center}
\end{figure*}

\section{Astrophysical implication}
\label{sec:Astro}
We examine the impact of our new $^{64}$Ge($p$,$\gamma$) and $^{65}$As($p$,$\gamma$) rates and their uncertainties on the $rp$-process using one-zone
XRB models. Post processing calculations using temperature and density trajectories from the literature enable a quick assessment of the impact of nuclear
physics changes on the strength of the $^{64}$Ge waiting point using the $A=64$ abundance, and on the burst energy generation rate. We use the post-processing
approach for the K04 X-ray burst model~\citep{par08,par09}. However, postprocessing calculations do not take into account the changes in temperature and density
that result from the energy generation changes. They can therefore not predict reliably the quantitative impact on produced abundances and light curves.
To account for this effect, we also use the full one-zone X-ray burst model~\citep{sch01}, which represents a more extreme burst with very hydrogen rich
ignition.

\subsection{Post processing results for K04}
With the representative K04 thermodynamic history~\citep{par08,par09}, final abundances (as mass fractions $X$) and the nuclear energy generation rate
$E_\mathrm{gen}$ during a burst have been studied by performing separate XRB model calculations with different rates. In the K04 model, the peak temperature
$T_\mathrm{peak}=1.4$~GK is similar to those reached at the base of the envelope in comparable hydrodynamic XRB models (e.g., 1.3 GK in~\citet{jos10}).
Figs.~\ref{fig3_Abund} and~\ref{fig4_Egen} compare results for $X$ and $E_\mathrm{gen}$ using rates from the present work to results using rates available in
JINA REACLIB:
\emph{laur, rath, rpsm, thra, ths8}. The impact of
(a) using different $^{64}$Ge($p$,$\gamma$) rates (with the $^{65}$As($p$,$\gamma$) rate held constant at the \emph{Present} value),
(b) using different $^{65}$As($p$,$\gamma$) rates (with the $^{64}$Ge($p$,$\gamma$) rate held constant at the \emph{Present} value) and
(c) using different $^{64}$Ge($p$,$\gamma$) and $^{65}$As($p$,$\gamma$) rates together, is indicated in each of the two figures.
For each change in reaction rate, the corresponding inverse reaction rate is also changed to maintain detailed balance. This inverse rate strongly
depends on the adopted reaction $S_p$ -value for the respective forward rate. As we compare the impact of different rates that have been determined using very
different $S_p$ values (see Sect. 2 and Figs. 1 and 2), the results illustrate not only the influence of the rate calculation, but also the influence of
different forward to reverse rate ratios due to different $S_p$-values.

The results of Figs.~\ref{fig3_Abund} and~\ref{fig4_Egen} are interesting, but not entirely unexpected. For the case of the $^{64}$Ge($p$,$\gamma$) reaction, an
equilibrium between the rates of the forward ($p$,$\gamma$) and reverse ($\gamma$,$p$) processes is quickly established due to the small ($p$,$\gamma$) $Q$-value
relative to $kT$ at XRB temperatures: at 1 GK, $kT \approx100$~keV. As a result, it is the ($p$,$\gamma$) $Q$-value, rather than the actual $^{64}$Ge($p$,$\gamma$)
rate, that is the most important nuclear physics quantity needed to characterize the equilibrium abundance of $^{65}$As (and the subsequent flow of material to
heavier nuclei through the $^{65}$As($p$,$\gamma$) reaction), c.f.~\citet{sch98,ili07,par09}. This is nicely illustrated in Fig.~\ref{fig3_Abund}(a): the
$^{64}$Ge($p$,$\gamma$) rates adopting positive ($p$,$\gamma$) $Q$-values (\emph{thra, laur, rath}) give relatively lower final abundances around $A = 64$ and
larger abundances at higher masses precisely because of the larger equilibrium abundances of $^{65}$As during the burst, allowing for increased flows of
abundances to higher masses via the $^{65}$As($p$,$\gamma$) reaction. On the other hand, the opposite is true for those rates adopting negative ($p$,$\gamma$)
$Q$-values (\emph{ths8, Present, rpsm}) because of the larger equilibrium abundances of $^{64}$Ge and lower relative abundances of $^{65}$As during the burst.
Indeed, the summed mass fractions of species with $A>70$, $X_{70}$, vary considerably for different choices of $Q$-values. For example, when the \emph{thra},
\emph{ths8} or \emph{Present} rates are adopted, $X_{70}$ = 0.58, 0.21, or 0.33, respectively.

A consequence of the increased flow of abundances to heavier nuclei is seen in Fig.~\ref{fig4_Egen}(a), where the models adopting the \emph{thra, laur}, and
\emph{rath} rates give the largest $E_\mathrm{gen}$ at late times due to energy released from the decay of the larger amounts of heavy nuclei produced during
the burst. As expected, the opposite is true for $E_\mathrm{gen}$ in the models using the \emph{ths8, Present} and \emph{rpsm} rates. We note that the predictions
for $E_\mathrm{gen}$ at late times vary rather significantly between the models using the different rates, with differences as large as a factor of $\approx$~2.

For the case of the $^{65}$As($p$,$\gamma$) reaction, where a large positive ($p$,$\gamma$) $Q$-value is adopted in all rate estimates, the importance of the
actual rate is illustrated in Fig.~\ref{fig3_Abund}(b). The model adopting the largest rate at the most relevant temperatures ($T > 0.9$ GK, c.f.~\citet{jos10}),
\emph{rath}, gives the lowest abundances around $A = 64$ and the largest abundances at higher masses. Again, as expected, the opposite is true for the model using
the lowest $^{65}$As($p$,$\gamma$) rate, \emph{laur}. The variation in $X_{70}$ for different choices of the $^{65}$As($p$,$\gamma$) rate is significant, with
$X_{70}$ = 0.49 with the \emph{rath} rate, and 0.29 with the \emph{laur} rate. The behavior of $E_\mathrm{gen}$ for these models is again in accord with the
distributions of the final abundances, with the largest $E_\mathrm{gen}$ at late times arising from the model using the \emph{rath} rate, and the lowest
$E_\mathrm{gen}$ arising from the model using the \emph{laur} rate.

Finally, the effects of using different $^{64}$Ge($p$,$\gamma$) and $^{65}$As($p$,$\gamma$) rates from the same theoretical model calculation are shown in Figs.~\ref{fig3_Abund}(c) and~\ref{fig4_Egen}(c). This reveals the impact of competing influences from these two rates. For example, the model using the \emph{ths8} $^{64}$Ge($p$,$\gamma$) rate gave the lowest relative abundances at higher masses (see Fig.~\ref{fig3_Abund}(a)), while the model using the \emph{ths8} $^{65}$As($p$,$\gamma$) rate gave among the highest relative abundances at higher masses (see Fig.~\ref{fig3_Abund}(b)). When these two rates are used together, the combined impact on the final abundances (and $E_\mathrm{gen}$) is, not surprisingly, moderated.

Fig.~\ref{fig3_Abund}(c) also shows that using the \emph{Present} $^{64}$Ge($p$,$\gamma$) and $^{65}$As($p$,$\gamma$) rates results in the strongest $^{64}$Ge waiting point, and the lowest final abundances at the highest masses. Differences with respect to predictions using rates from JINA REACLIB are as large as a factor of $\approx$~7 at individual values of $A$. $X_{70}$ calculated with the two \emph{Present} rates differs by as much as factor of 1.8 from models using the other rates.

We have also examined the impact on XRB model predictions of our uncertainties in the \emph{Present} $^{64}$Ge($p$,$\gamma$) and $^{65}$As($p$,$\gamma$) rates, as shown in Figs.~\ref{Fig1_64Gep} and~\ref{Fig2_65Asp}. Reverse rates for the lower and upper forward rates were determined using exactly the $Q$-values adopted for the corresponding forward rate calculations. Panels (a) and (b) of Figs.~\ref{fig3_Abund} and~\ref{fig4_Egen} show how these rate uncertainties affect final abundances and $E_\mathrm{gen}$ in the K04 model, with mass fractions above $A = 64$ varying by up to a factor of 3 due to the individual uncertainties in the rates, $X_{70}$ varying by up to a factor of 2, and $E_\mathrm{gen}$ varying by up to 35\% at late times. The impact on $X$ and $E_\mathrm{gen}$ of the uncertainties in the \emph{Present} rates is somewhat smaller than that from different choices of rates but clearly not insignificant. As such, the mass of $^{66}$Se should be determined experimentally and the uncertainty in the mass of $^{65}$As should be reduced to better
constrain model predictions.

\begin{figure}[t]
\begin{center}
\hspace{-20mm}
\includegraphics[scale=0.35,angle=-90]{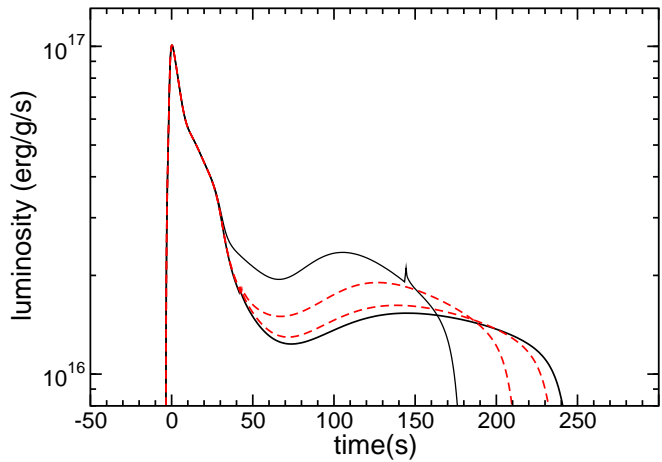}
\vspace{-15mm}
\caption{Luminosity per gram of material calculated with the one-zone X-ray burst model for nuclear physics input that, within uncertainties, maximally favours
(initially high solid line) and maximally disfavours (initially low solid line) the $rp$-process flow through the $^{64}$Ge waiting point. The dashed lines show
the results when only the $^{65}$As($p$,$\gamma$) reaction rate is changed within uncertainties, and all other nuclear physics input is fixed. (A color version
of this figure is available in the online journal.)}
\label{fig5_light}
\end{center}
\end{figure}

\begin{figure}[t]
\begin{center}
\hspace{-20mm}
\includegraphics[scale=0.35,angle=-90]{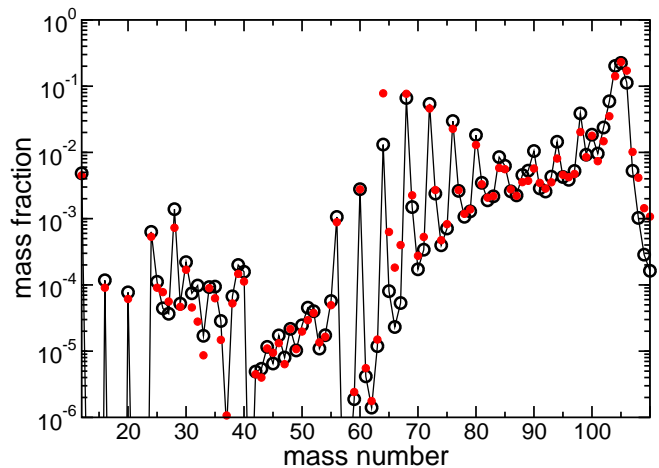}
\vspace{-15mm}
\caption{Final mass fractions, summed by mass number, calculated with the one-zone X-ray burst model for nuclear physics input that, within uncertainties,
maximally favours (open black circles, solid line) and maximally disfavours (filled red circles) the $rp$-process flow through the $^{64}$Ge waiting point.
(A color version of this figure is available in the online journal.)}
\label{fig6_abundances}
\end{center}
\end{figure}

\begin{figure}[t]
\begin{center}
\hspace{-20mm}
\includegraphics[scale=0.35,angle=-90]{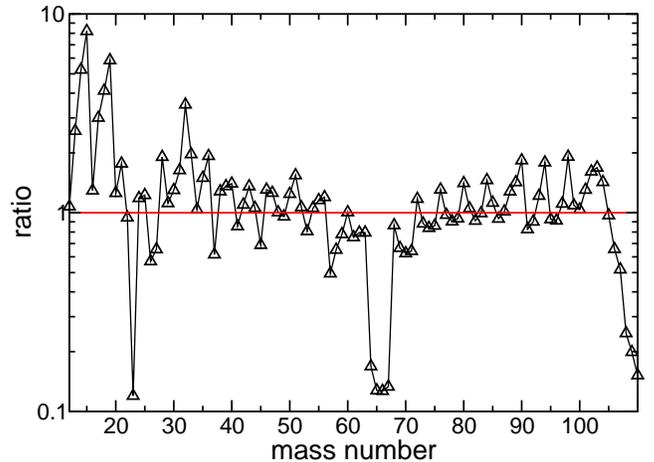}
\vspace{-15mm}
\caption{Ratio of the final mass fractions shown in Fig.~\ref{fig6_abundances}.}
\label{fig7_ratio}
\end{center}
\end{figure}

\subsection{One-zone X-ray burst model results}

In addition we explored the impact of the remaining nuclear physics uncertainties related to the $^{64}$Ge waiting point on the full one-zone X-ray burst model described in~\citet{sch01}. We note that this model is different from K04. It represents a burst that ignites in a very hydrogen rich environment, for example at high accretion rates and very low accreted metallicity, and was developed to explore the maximum extent of the $rp$-process towards heavy elements. To determine the total remaining uncertainty in the burst model due to the nuclear physics of the $^{64}$Ge waiting point, we performed two extreme calculations. The calculations assume the most favourable (unfavourable) nuclear physics choice for the $rp$-process to pass through the $^{64}$Ge waiting point, adopting the upper (lower) limit of the $^{64}$Ge($p$,$\gamma$) and $^{65}$As($p$,$\gamma$) reaction rates, and the upper (lower) limits of $S_{\rm p}(^{65}$As) and $S_{\rm p}(^{66}$Se). Varying $S_{\rm p}$ values independently, rather than varying individual masses, is justified as the uncertainties in $S_{\rm p}(^{65}$As) and $S_{\rm p}(^{66}$Se) are each completely dominated by the mass uncertainty of $^{65}$As and $^{66}$Se, respectively. Figs.~\ref{fig5_light} and~\ref{fig6_abundances} show the impact of the nuclear physics uncertainties on burst light curve and final composition. Clearly the nuclear physics uncertainties have a strong impact on observables. The abundance ratio of $A=64$ to $A=68$, a measure for the strength of the $^{64}$Ge waiting point varies from 1.1 (making $^{64}$Ge the strongest waiting point) to 0.2 (making $^{64}$Ge not a significant waiting point). This is consistent with the result from~\citet{tu11} who found that with their new $^{65}$As mass $^{64}$Ge is only a weak waiting point. However, as we show here, when taking into account all nuclear physics uncertainties, a strong $^{64}$Ge waiting point cannot be ruled out. Fig.~\ref{fig7_ratio} shows the ratio of the final abundances. Similar to the results obtained for model K04 using post-processing, a weak $^{64}$Ge waiting point (low $^{64}$Ge abundance) leads to an enhancement of the production of heavier elements by up to a factor of 2. As already noted by~\citet{tu11} the production of the heaviest nuclei with $A \ge 106$ is however reduced for a weak $^{64}$Ge waiting point. This somewhat counter intuitive result is a consequence of the faster burning at higher temperature, which leads to a shorter burst as hydrogen is consumed more quickly. This effect is not seen in the post-processing calculation because there the temperature trajectory is fixed.

We also investigated the relative contributions of the various nuclear physics uncertainties. Similar to the K04 post processing results we find that the $^{64}$Ge($p$,$\gamma$) reaction rate itself, for a fixed $Q$-value, has no influence on the burst model due to ($p$,$\gamma$)-($\gamma$,$p$) equilibrium between $^{64}$Ge and $^{65}$As. A calculation where only $S_{\rm p}(^{65}$As) and $S_{\rm p}(^{66}$Se) changes produced virtually the same result as the full variation, demonstrating that the mass uncertainties currently dominate. Changing each $S_{\rm p}$ separately indicates that the 85~keV uncertainty of $S_{\rm p}(^{65}$As) due to the $^{65}$As mass uncertainty, and the 310 keV uncertainty of $S_{\rm p}(^{66}$Se) mainly due to the unknown $^{66}$Se mass contribute roughly equally. However, varying the $^{65}$As($p$,$\gamma$) reaction rate within our new uncertainties, while leaving $S_{\rm p}(^{65}$As) and $S_{\rm p}(^{66}$Se) fixed at their nominal values, still led to significant light curve changes (see Fig.~\ref{fig5_light}) and a change of the $A=64$ to $A=68$ ratio from 1 to 0.7. This shows that once the mass uncertainties are addressed, the $^{65}$As($p$,$\gamma$) reaction rate uncertainty will still play a role (even though fixing $S_{\rm p}(^{66}$Se) will reduce the rate uncertainty somewhat).

\section{Summary and conclusion}

We have determined new thermonuclear rates for the $^{64}$Ge($p$,$\gamma$)$^{65}$As and $^{65}$As($p$,$\gamma$)$^{66}$Se reactions based on large-scale shell model calculations and proton separation energies for $^{65}$As and $^{66}$Se derived using measured masses and, for $^{66}$Se, the AME2012 extrapolation. These rates differ strongly from other rates available in the literature. For example, at $\approx$~1 GK, our $^{64}$Ge($p$,$\gamma$) rate is up to a factor of $\approx$~90 lower than other rates, while our $^{65}$As($p$,$\gamma$) rate differs by up to a factor of $\approx$~3 from other rates.

We also determined for the first time reliable uncertainties for the $^{64}$Ge($p$,$\gamma$)$^{65}$As and $^{65}$As($p$,$\gamma$)$^{66}$Se reactions. We find that in two different X-ray burst models, the remaining uncertainties in $S_{\rm p}(^{65}$As), $S_{\rm p}(^{66}$Se), and the $^{65}$As($p$,$\gamma$) reaction rate lead to large uncertainties in the strength of the $^{64}$Ge waiting point in the $rp$-process, the produced amount of $A=64$ material in the burst ashes that will ultimately decay to $^{64}$Zn, the produced amount of heavier nuclei beyond $A=64$, and the burst light curve. These effects are robust and appear in two different X-ray burst models. We conclude that to address these uncertainties a more precise measurement of the $^{65}$As mass, a measurement of the $^{66}$Se mass, and a measurement of the excitation energies of states in $^{66}$Se that serve as important resonances for the $^{65}$As($p$,$\gamma$)$^{66}$Se reaction will be important.

\acknowledgments

This work was financially supported by the Major State Basic Research Development Program of China (2013CB834406), and the National Natural Science Foundation of China (Nos. 11490562, 11135005, 11321064, U1432125, U1232208). YHL gratefully acknowledges the financial supports from Ministry of Science and Technology of China (Talented Young Scientist Program) and from the China Postdoctoral Science Foundation (2014M562481). AP was supported by the Spanish MICINN under Grant No. AYA2013-42762. HS was supported by the US National Science Foundation under Grant No. PHY-1430152 (JINA Center for the Evolution of the Elements) and PHY 11-02511. BAB was supported by the US National Science Foundation under Grant No. PHY-1404442.

\end{document}